\documentclass[preprint,10pt]{article}
\usepackage{multicol}
\usepackage[T1]{fontenc}
\usepackage{amsmath}
\usepackage{amssymb}
\usepackage{natbib}
\usepackage{hyperref}
\usepackage{xcolor}
\definecolor{lightgray}{gray}{0.90}
\usepackage{graphicx}
\usepackage{subcaption}
\usepackage{caption}
\usepackage{geometry}
\begin{document}

\begin{minipage}[t]{\textwidth}
  \centering
  {\LARGE IceCube PeV neutrinos from heavy dark matter decay with 12 years HESE data}\\[2ex]
  {\large
    Diptarko Mukherjee\textsuperscript{a}\footnotemark[1],
    Ashadul Halder\textsuperscript{b}\footnotemark[2],
    Debasish Majumdar\textsuperscript{a}\footnotemark[3],
    Abhijit Bandyopadhyay\textsuperscript{a}\footnotemark[4]
  }\\[1ex]
  \textsuperscript{a}\textit{Department of Physics, Ramakrishna Mission Vivekananda Educational and Research Institute, Belur Math, \\Howrah 711202, India}\\
  \textsuperscript{b}\textit{Department of Physics, St. Xavier’s College, Kolkata 700016, India}
\end{minipage}
\footnotetext[1]{Email: dpcybertron2001@gmail.com}
\footnotetext[2]{Email: ashadul.halder@gmail.com}
\footnotetext[3]{Email: debasish.majumdar@gm.rkmvu.ac.in}
\footnotetext[4]{Email: abhijit.phy@gm.rkmvu.ac.in}

\vspace{0.5cm}
\begin{center}
    \begin{minipage}[t]{0.80\textwidth}
    \begin{center}
      \textbf{Abstract}\\[-0.5ex]
    \end{center}
      \vspace{0.2cm}
      The decay of superheavy dark matter from the early universe may undergo decay via QCD cascades and electroweak cascade to produce neutrinos as one of the decay products. We consider the neutrino events in and around PeV region reported by IceCube collaboration are due to the decay of such heavy dark matter. The neutrino spectrum could be from the decay processes via hadronic decay modes and/or leptonic decay modes. Using the numerical evolution of QCD cascades as well as electroweak corrections where use has been made of DGLAP equations, the neutrino fluxes from the heavy dark matter decay have been computed. The mass of the decaying superheavy dark matter and its decay lifetime have then been estimated from a $\chi^2$ analysis of the IceCube 12-year data. The fractional contribution ($f_{\rm lep}$) of the leptonic decay channel in such a decay process is also estimated from the same $\chi^2$ analyses. It is seen that to explain the IceCube 12-year ultrahigh energy (UHE) events the mass of a decaying superheavy dark matter would be $\sim9.4\times 10^6$ GeV and decay time $\tau \simeq 4.2 \times 10^{28}$ second. It is also found that the lepton channel contribution is very small, $f_{\rm lep} \sim 0.001$.
    \end{minipage}
\end{center}

\vspace{2ex}
\rule{\textwidth}{0.5pt}
\vspace{1ex}


\section{Introduction}

After the observation of the first PeV neutrino more than a decade back at IceCube (IC), enormous number of data have been collected. Out of these the High Energy Starting Event (HESE) data have been analysed extensively. Recently, the IceCube collaboration has published \cite{IceCube:2023sov} twelve year data and then have performed a thorough analysis with the HESE data samples of this twelve year data set. As reported by IceCube collaboration 64 HESE new events are considered from HESE data and these data are combined with previous 102 events \cite{PhysRevD.104.022002}. Therefore, this analysis has been performed with total 164 HESE events. They have omitted event number 128 and 132 since those contain coincident atmospheric background. With these 164 HESE events, IceCube collaboration provided the most up-to-date per event reconstructed quantities. In doing this, they have used more refined techniques toward addressing physics parameter space, ice models etc. From their analyses, they have found a muon track event with most probable visible energy to be 4.8 PeV. From this track the initial muon energy is estimated to be ($9\pm 4$) PeV \cite{Chirkin:2004hz}, which corresponds to a total energy of about $13\pm 5$ PeV \cite{Chirkin:2004hz}.

Such ultra high energy (UHE) events could be originated from very high energy processes in astrophysical bodies such as active galactic nuclei (AGN), GRBs, blazars or the likes. But in this work we consider the origin of UHE neutrinos via a different mechanism whereby a super-heavy dark matter at the GUT scale decays to produce UHE neutrinos of energies in UHE range (hundreds of TeV and beyond). Such super-heavy dark matter could have been produced in early Universe by the process of gravitational production mechanism \cite{Chung_2001,gelmini}.The decay of such a super-heavy dark matter with masses larger than electroweak scale proceeds via QCD cascades \cite{Berezinsky:1997sb,Berezinsky:1999yk}. It may also undergo electroweak cascade. A decay via QCD cascade may proceed via two channels of which one is hadronic channel whereby the heavy particle $\chi$ decays to $\bar{q}q$ ($q$ being a quark with any flavour) and the other is via leptonic channel. For the hadronic channel the hadrons are the decay products ultimately that decays down to $\gamma$, $\nu$, $e$ etc. In leptonic decay channel however, the decay process $\chi \rightarrow$ $\mu$ , $\nu$ would be sufficient in the generic case.

In this work we consider the possible signals of ultra high energy neutrinos detected by IceCube to have originated from the decay of super-heavy dark matter discussed above. We use the latest 12-year HESE data published by IceCube \cite{IceCube:2023sov} and analyse this data to estimate the mass and decay lifetime of such a primordial super-heavy dark matter. In our analysis we have considered both the hadronic and leptonic channels of the super-heavy dark matter decay.

The paper is organized as follows. In section~\ref{section2}, we provide a brief account for the formalism and the neutrino spectrum obtained from expressions of the neutrino spectra for the two decay channel namely hadronic cascade channel and the leptonic channel. In section~\ref{section3}, we furnish calculations and the results. Finally, in section~\ref{section4} contains some discussions and concluding remarks.

\section{\label{section2} Decay of Super-heavy dark matter}

The decay of super-heavy dark matter (SHDM) particles, which may have originated in the early Universe, involves a cascade of QCD partons. When the mass of the SHDM particle $m_{\chi}$ is significantly greater than the electroweak scale ($m_{\chi} \gg m_W$, $m_W$ is the mass of $W$ boson), both QCD and electroweak cascades contribute to the decay process~\cite{bere1,bere2}. The hadronic spectrum produced through QCD cascades including possible contributions from supersymmetric QCD can be evaluated using two primary methods. The first is based on Monte Carlo (MC) simulations~\cite{bere3,bere4} and the second involves the Dokshitzer-Gribov-Lipatov-Altarelli-Parisi (DGLAP) equations~\cite{bere4}--\cite{cyrille2}, which describe the evolution of fragmentation functions. In this work, we use the neutrino spectrum or flux obtained from both the DGLAP evolution and MC approaches to interpret the IceCube events detected at UHE range. The generation of the neutrino spectrum is attributed to two main decay channels namely hadronic and leptonic.

\subsection{Hadronic Decay Channels}

The decay of superheavy particles can lead to the production of hadrons through the process of QCD cascade. In spite of smaller QCD coupling, a cascade still appears as parton splitting becomes more frequent due to the presence of large logarithmic enhancements associated with soft parton emissions. Electroweak radiative corrections can also be influenced by similar logarithmic effects. To study the QCD cascade, we use a numerical code that solves the DGLAP evolution equations, as described in \cite{bere4}.

In this section, we examine the hadronic decay channel $\chi \rightarrow q \bar{q}$. Following the decay process, the quarks undergo a perturbative QCD cascade, which results in hadronization. The produced hadrons are mostly unstable and subsequently decay into leptons and other final-state particles. Compared to other theoretical uncertainties, the impact of electroweak radiative corrections on the development of the cascade is relatively small.

The final neutrino spectrum is written as
\begin{equation}
\frac{dN_\nu}{dx} = 2R \int_{xR}^{1} \frac{dy}{y} \, D^{\pi^\pm}(y) + 2 \int_{x}^{1} \frac{dz}{z} \, f_{\nu_i}\left( \frac{y}{z} \right) D^{\pi^\pm}(z), \label{eq:nu_flx}
\end{equation}

where, the factor $R = 1/(1 - r)$, where $r = \left( m_{\mu}/m_{\pi} \right)^2 \approx 0.573$. In the above equation (Eq.~\ref{eq:nu_flx}), the fragmentation function of the pions $D_{\pi}(x, s)$ from a parton $i =(\text{quarks},\,\text{gluons})$ is defined as 
\begin{equation}
D_{\pi}(x, s) \equiv D_{\pi q}(x, s) + D_{g \pi}(x, s),
\end{equation}
where $x \equiv 2E/m_\chi$ is a dimensionless quantity, representing the energy of the super-heavy dark matter transferred to the hadron. 

The functions $f_{\nu_i}(x)$ are defined as follows, which are adopted from Ref.~\cite{kelner}

\begin{equation}
    \begin{aligned}
        f_{\nu_i}(x) &= g_{\nu_i}(x)\,\Theta(x - r) + \left(h_{\nu_i}^{(1)}(x) + h_{\nu_i}^{(2)}(x)\right)\,\Theta(r - x), \\
        g_{\nu_\mu}(x) &= \frac{3 - 2r}{9(1 - r)^2} \left(9x^2 - 6\ln x - 4x^3 - 5\right), \\
        h_{\nu_\mu}^{(1)}(x) &= \frac{3 - 2r}{9(1 - r)^2} \left(9r^2 - 6\ln r - 4r^3 - 5\right), \\
        h_{\nu_\mu}^{(2)}(x) &= \frac{(1 + 2r)(r - x)}{9r^2} \left[9(r + x) - 4(r^2 + rx + x^2)\right], \\
        g_{\nu_e}(x) &= \frac{2}{3(1 - r)^2} \left[(1 - x)\left(6(1 - x)^2 + r(5 + 5x - 4x^2)\right)+ 6r \ln x\right], \\
        h_{\nu_e}^{(1)}(x) &= \frac{2}{3(1 - r)^2} \left[(1 - r)(6 - 7r + 11r^2 - 4r^3) + 6r \ln r\right], \\
        h_{\nu_e}^{(2)}(x) &= \frac{2(r - x)}{3r^2} \left(7r^2 - 4r^3 + 7xr - 4xr^2 - 2x^2 - 4x^2r\right).\nonumber
    \end{aligned}
\end{equation}
Here, $\Theta$ denotes the Heaviside step function, defined as $\Theta(x) = 1$ for $x \geq 0$ and $\Theta(x) = 0$ for $x < 0$.

Fragmentation functions at high energy scales can be determined by evolving the Dokshitzer–Gribov–Lipatov–Altarelli–Parisi (DGLAP) evolution equations \cite{Gribov:1972rt,ALTARELLI1977298,Dokshitzer:1977sg} starting from experimental values at low scale. Detailed discussions on this procedure can be found in Refs.~\cite{Kalashev_2016,bere4}. In our analysis, we employ the pion fragmentation functions, where they are averaged over different quark flavors and interpolated down to values of $x$ as low as $10^{-5}$. Since pion decays are the primary contributors to the resulting neutrino flux, other mesonic contributions are not included in our analysis, as their impact has been estimated to be $\lesssim 10 \%$ \cite{bere4}.

\subsection{Leptonic Decay Channels}

The electroweak cascade process can be explored by considering the tree-level decay of a super-heavy particle of mass $m_{\chi} \leq m_{\rm GUT}$. According to the framework of the $Z$-burst model, the decay predominantly proceeds via $\chi \rightarrow \bar{\nu} \nu$ channel, though it may decay into charged lepton pairs as well. When $m_{\chi}$ is much larger than the $Z$ boson mass ($m_Z$) and considering a typical momentum transfer bounded by $Q^2 \leq m_{\chi}^2 / 4$, the $Z$ boson mass becomes negligible in the kinematic analysis. Although the QCD coupling remains small at high energies, its effects become significant due to large logarithmic factors of the form $\ln^2(m_{\chi}^2 / m_Z^2)$, which arise from soft and collinear emissions.

A similar enhancement occurs when $m_{\chi} \gg m_W$, due to the presence of large logarithmic corrections that make fixed-order perturbation theory unfeasible. This results in the formation of an electroweak cascade, closely analogous to QCD cascade. Moreover, interactions between QCD and electroweak cascades can also take place, as electroweak bosons like the $W$ and $Z$ may decay into quark pairs, resulting in slight modification of the final state hadronic spectrum. At the same time, processes such as $W \rightarrow \bar{\nu} \nu$ contributes to additional neutrinos through the electroweak sector.

In order to study the neutrino spectrum from decay processes in leptonic decay modes, comprehensive Monte Carlo simulations have been performed. These include both QCD-induced cascades~\cite{bere3} and electroweak showering effects~\cite{bere6}. The hadronization process is adopted from Ref.~\cite{bere4}.

\begin{figure}
    \centering
    \begin{subfigure}[b]{0.49\textwidth}
        \includegraphics[trim=0 45 0 45, clip, width=\textwidth]{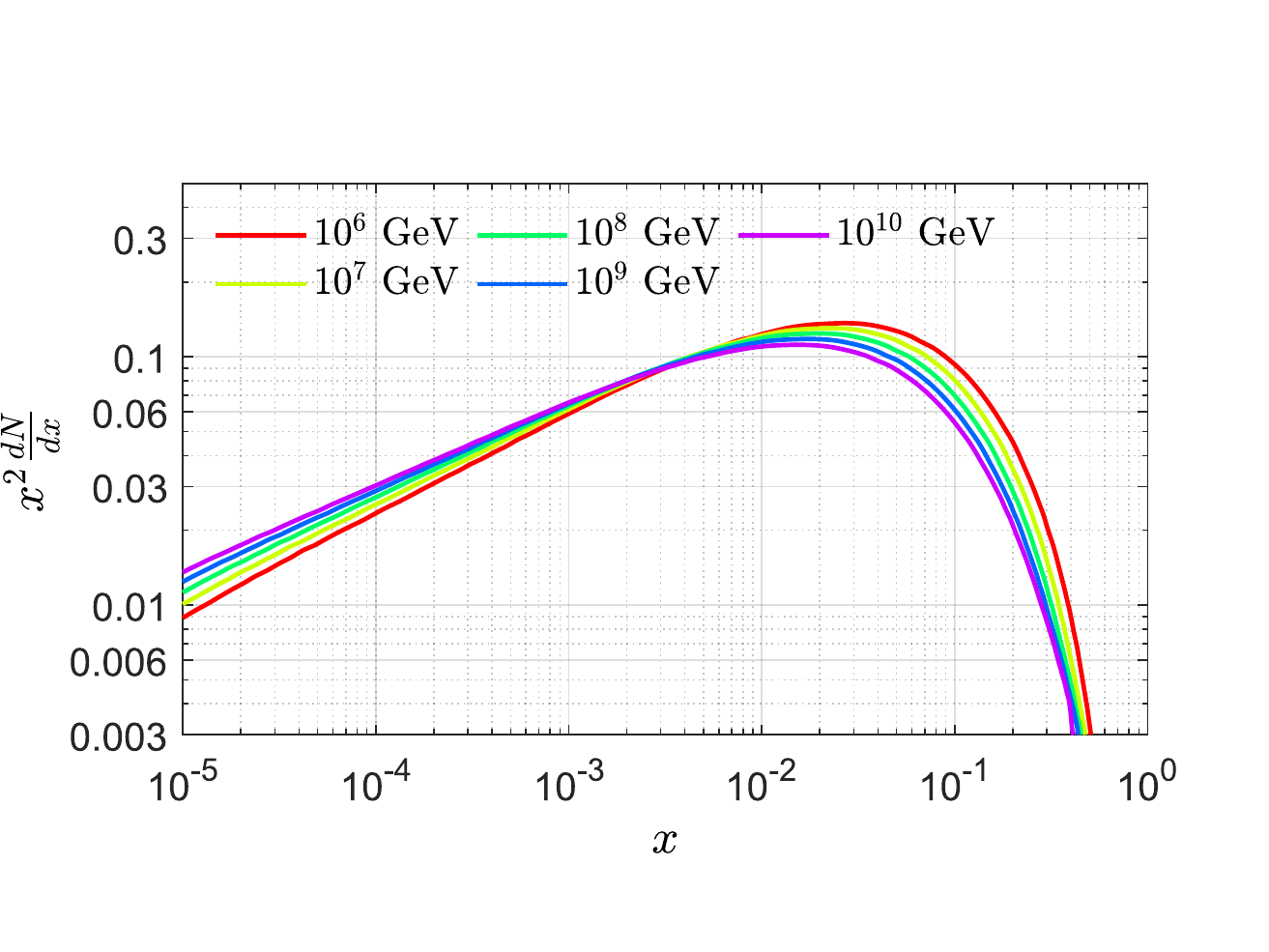}
        \caption{Hadronic neutrino flux}
        \label{fig:hadronic}
    \end{subfigure}
    \hfill
    \begin{subfigure}[b]{0.49\textwidth}
        \includegraphics[trim=0 45 0 45, clip, width=\textwidth]{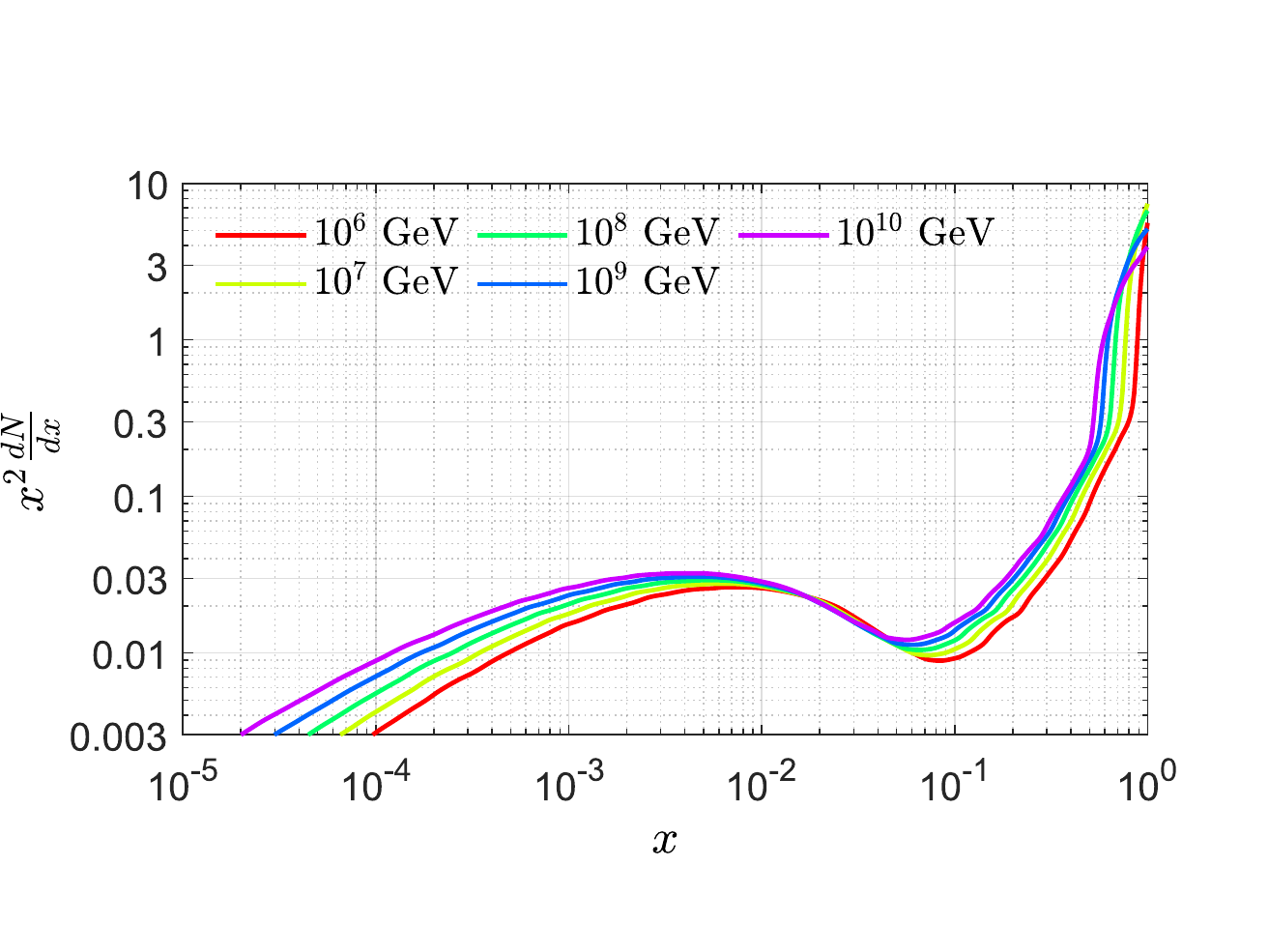}
        \caption{Leptonic neutrino flux}
        \label{fig:leptonic}
    \end{subfigure}
    \caption{Comparison of neutrino fluxes for different $m_{\chi}$}
    \label{fig:flux_comparison}
\end{figure}
Fig.~\ref{fig:flux_comparison} shows a comparative analysis of the hadronic and leptonic neutrino fluxes as functions of the dimensionless energy fraction $x$ for five logarithmically spaced values of the dark matter particle mass $m_{\chi}$ ranging from $10^6$~GeV to $10^{10}$~GeV. In the left panel, the spectrum originated from hadronic fluxes are shown. Conversely, the right panel displays the leptonic flux, which tends to be softer especially at lower $x$.

\subsection{Neutrino flux}

The neutrino flux produced by the decay of super-heavy dark matter consists of two parts- one from within our galaxy (the Galactic component) and the other from outside of our galaxy (the extragalactic component). The galactic neutrino flux resulting from the decay of super-heavy dark matter with mass $m_{\chi}$ and decay lifetime $\tau$ can be written as 
\begin{equation}
    \dfrac {d\Phi_{\rm G}}{dE_\nu} (E_\nu) = \dfrac{1}{4\pi m_\chi \tau} \int_{V} \dfrac{\rho_{\chi}  (R[r])}{4\pi r^2} \dfrac{dN}{dE} (E,l,b)dV.
    \label{eq:gal}
\end{equation}

In the above expression, $\frac{dN}{dE}(E,l,b)$ denotes the neutrino spectrum arising from the decay of super-heavy dark matter and $l$($b$) are the Galactic longitude(latitude) respectively. The quantity $\rho_{\chi} (R[r])$ is the dark matter density, which is a function of the distance $R$ from the Galactic Center and $r$ denotes the distance from the Earth. In this work, we employ the Navarro-Frenk-White (NFW) profile to model the dark matter distribution \cite{nfw1,nfw2}. The integration is performed over the Milky Way halo with the maximum distance from the Galactic Center set to $R_{\rm max} = 260$~kpc, as suggested in Ref.~\cite{milky}.

The extragalctic component of the neutrino flux originated from similar decay can be expressed as
\begin{equation} 
    \frac{d\Phi_E}{dE}(E_\nu) = \frac{1}{4\pi m_\chi \tau} \int_0^{\infty} \frac{\rho_0 c}{H_0 \sqrt{\Omega_m (1 + z)^3 + (1 - \Omega_m)}} \, \frac{dN}{dE}\left[E(1 + z)\right]\,dz
    \label{eq:exgal}
\end{equation}
where $z$ is the cosmological redshift and $E(z)=(1+z)E$. In the above equation (Eq.~\ref{eq:exgal}), the proper radius of the Hubble sphere is $c/H_0 = 1.37 \times 10^{28}$ cm and the quantity $\rho_0 = 1.15 \times 10^{-6}~\text{GeV/cm}^3$ indicates the average cosmological dark matter density at the present epoch (redshift $z = 0$). The parameter $\Omega_m = 0.316$ denotes the present value of cosmological matter density parameter.

In the present analysis we introduce a parameter $f_{\rm lep}$ which indicates the fraction of super-heavy dark matter that decay through leptonic decay channel, while the rest (i.e. $1-f_{\rm lep}$) undergoes hadronic decay channel. The total differential flux obtained from both Galactic and extragalactic origins is given by the expression
\begin{equation}
    \dfrac{d\Phi_{\text{th}}}{dE}(E_\nu) = (1-f_{\rm lep})\left[\dfrac{d\Phi_{\text{EG}}}{dE}(E_\nu) + \dfrac{d\Phi_{\text{G}}}{dE}(E_\nu)\right]_{\rm hadronic}+f_{\rm lep}\left[\dfrac{d\Phi_{\text{EG}}}{dE}(E_\nu) + \dfrac{d\Phi_{\text{G}}}{dE}(E_\nu)\right]_{\rm leptonic}.
    \label{eq:total_nu}
\end{equation}

In this analysis, we assume that neutrino oscillations result in an equal flavour ratio of the neutrino flux upon arrival at Earth, i.e., $\nu_e:\nu_\mu:\nu_\tau = 1:1:1$. Under this assumption, we compute the total $\nu_{\mu}$ flux expected to reach the IceCube detector and compare it with the latest 12-year High-Energy Starting Events (HESE) dataset published by the IceCube Collaboration~\cite{IceCube:2023sov}. From this comparison, the best-fit values for the mass $m_{\chi}$ and the decay lifetime $\tau$ of the super-heavy dark matter particles can be estimated.

\section{\label{section3} Calculations and Results}

The purpose of this analysis is to estimate the mass and lifetime of a heavy dark matter, in case the UHE neutrino events detected by IceCube originate from the decay of such a super heavy dark matter. To this end we made a $\chi^2$-analysis of the 12-year HESE data~\cite{IceCube:2023sov}. The theoretical neutrino flux, as defined in Equation~\eqref{eq:total_nu} is treated as the model prediction in our analysis.      

The chi-square is defined as,
\begin{equation}
    \chi^2 = \sum_{i = 1}^{n} \left( \frac{E_i^2 \Phi_{{\rm th},i} - E_i^2 \Phi_{{\rm ex},i}}{\text{err}_i} \right)^2, \label{eq:chi2}
\end{equation}
where $E_i^2 \Phi_{{\rm th},i}$ and $E_i^2 \Phi_{{\rm ex},i}$ denote the theoretical and experimental flux values in the $i$-th energy bin, and $\text{err}_i$ is the corresponding uncertainty.
The Eq~ \ref{eq:chi2} is now minimized by varying three unknown parameters namely $m_\chi$, $\tau$, $f_{\rm lep}$ and the best fit values of these parameters are obtained.In addition the $\chi^2_{\rm min}$ results are also obtained for any two pair of parameters (out of the three parameters considered) by marginalizing over the third.

The IceCube data as obtained are in the form of a list of detected events characterized by their deposited energy without a direct flux measurement. In order to reconstruct the neutrino flux from this dataset, we utilize a Markov Chain Monte Carlo (MCMC) method. In this process, we incorporate various detector specifications such as the effective area of the IceCube detector, exposure time, and energy resolution. These inputs are essential for accurately modelling the detector response and for obtaining a reliable estimate of the flux, which is then used for comparison with the theoretical prediction through the chi-square analysis.

\begin{figure}
    \centering{}
    \includegraphics[width=\linewidth]{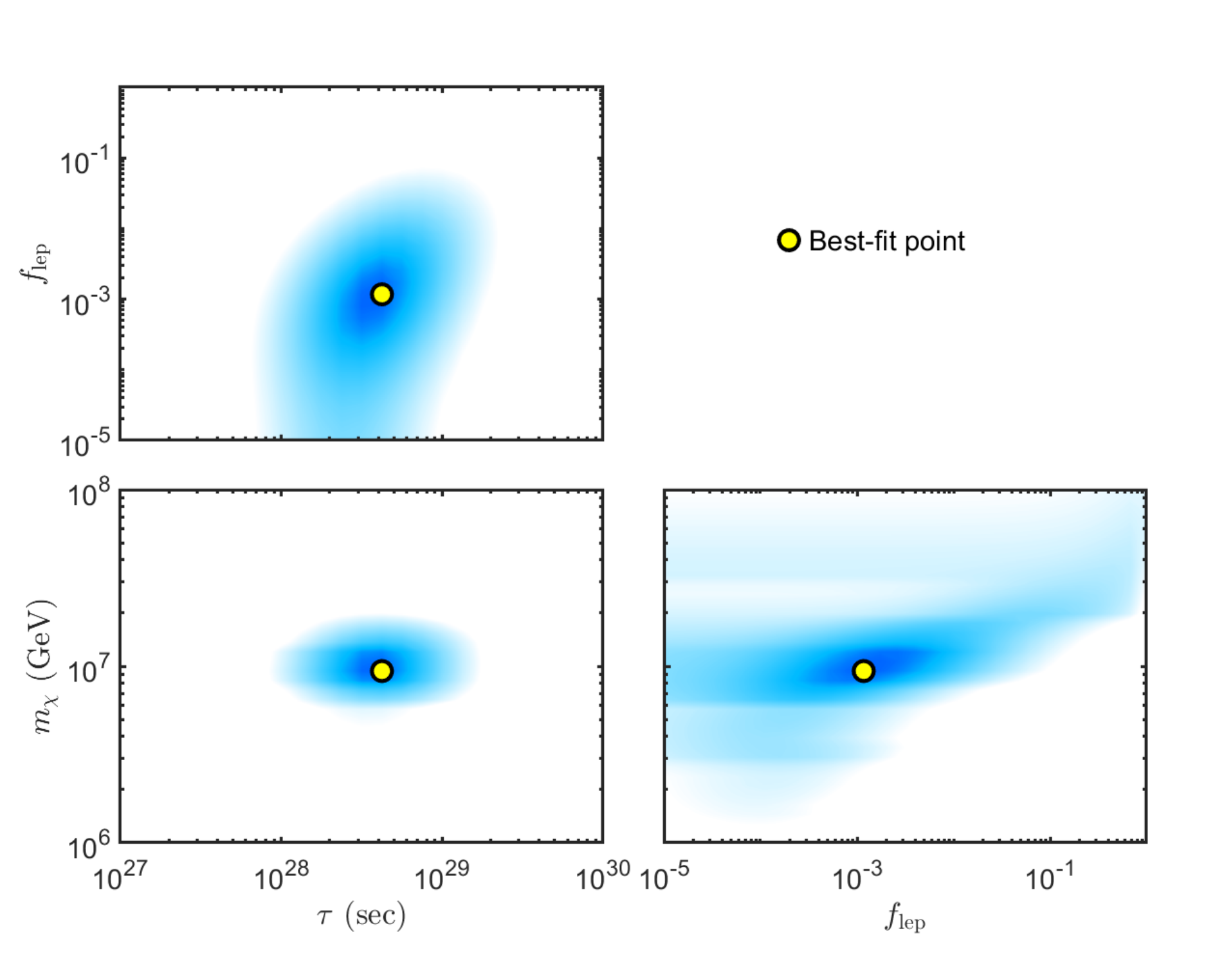}
    \caption{\label{fig:chi2} Two-dimensional projections of the chi-square fitting results in the three-dimensional parameter space ($m_{\chi} - \tau - f_{\rm lep}$ space). Each panel shows the $\chi^2$ values in each pair of parameter plains, marginalized over the third. The blue shaded areas represent lower $\chi^2$ values (better fit), with darker shades indicating closer agreement with IceCube data. The yellow dot with a black border marks the best-fit point in the full 3D parameter space.}
\end{figure}

The $\chi^2$ analyses of the data (mentioned above) are performed using (Eqs.~\ref{eq:gal} - ~\ref{eq:chi2}). In this case there are three unknown parameters namely super-heavy dark matter mass $m_\chi$, decay lifetime $\tau$, and leptonic branching ratio $f_{\rm lep}$. In   Fig.~\ref{fig:chi2} we present the results of the chi-square minimization described above. In this figure we show the best fit and the 1-$\sigma$, 2-$\sigma$, 3-$\sigma$, (C.L) contours or regions for each of the three pairs of parameters namely $(\tau, m_\chi)$, $(f_{\rm lep}, m_\chi)$, and $(f_{\rm lep}, \tau)$ out of the three parameters considered here. Therefore, each of the three panels of figure 2 Fig.~\ref{fig:chi2} is a two-dimensional projection for each pair of parameters.

Each of the three panel shows the $\chi^2_{\rm min}$ values and the allowed regions for a pair of parameters. Darker blue regions indicate lower $\chi^2$ values, reflecting better agreement with the observed IceCube 12-year HESE data. The best-fit point in the full three-dimensional space is shown as a, isolated yellow dot in Fig.~\ref{fig:chi2}  with a thick black outline. The best-fit values in this analysis are obtained as $m_\chi = 9.40 \times 10^6~\mathrm{GeV}$, $\tau = 4.22 \times 10^{28}~\mathrm{sec}$ and $f_{\rm lep} = 0.001$. It appears from this analysis that the leptonic channel contribution to the neutrino flux is very small ($10^{-3}$).

\begin{figure}
    \centering{}
    \includegraphics[width=0.7\linewidth]{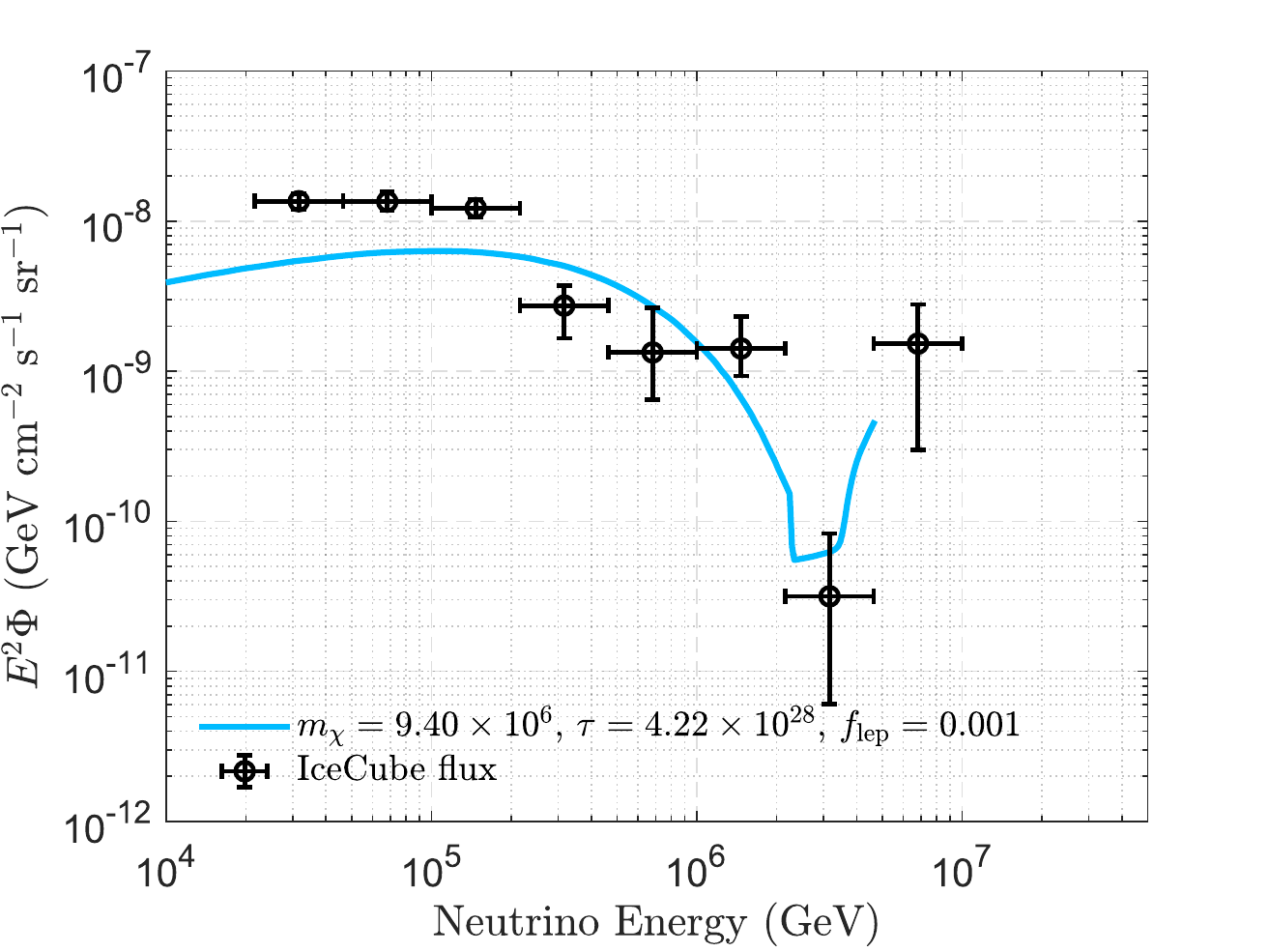}
    \caption{\label{fig:best_flux} Comparison between the neutrino flux from super-heavy dark matter decay with the best-fitted parameter values (solid blue curve) and the observed IceCube 12-year HESE data (black data points with error bars).}
\end{figure}

Fig.~\ref{fig:best_flux} illustrates the comparison between the theoretically computed neutrino flux from the decay of super-heavy dark matter and the flux reconstructed from the IceCube 12-year High-Energy Starting Events (HESE) data. The black circular points represent the IceCube data, showing the differential neutrino flux across different energy bins with associated statistical uncertainties.

The solid blue curve shows the theoretical flux prediction obtained by minimizing the $\chi^2$ (Eq~ \ref{eq:chi2}) by varying the three parameters namely $m_\chi$, $\tau$ and $f_{\rm lep}$. 

\begin{figure}
    \centering{}
    \includegraphics[width=0.7\linewidth]{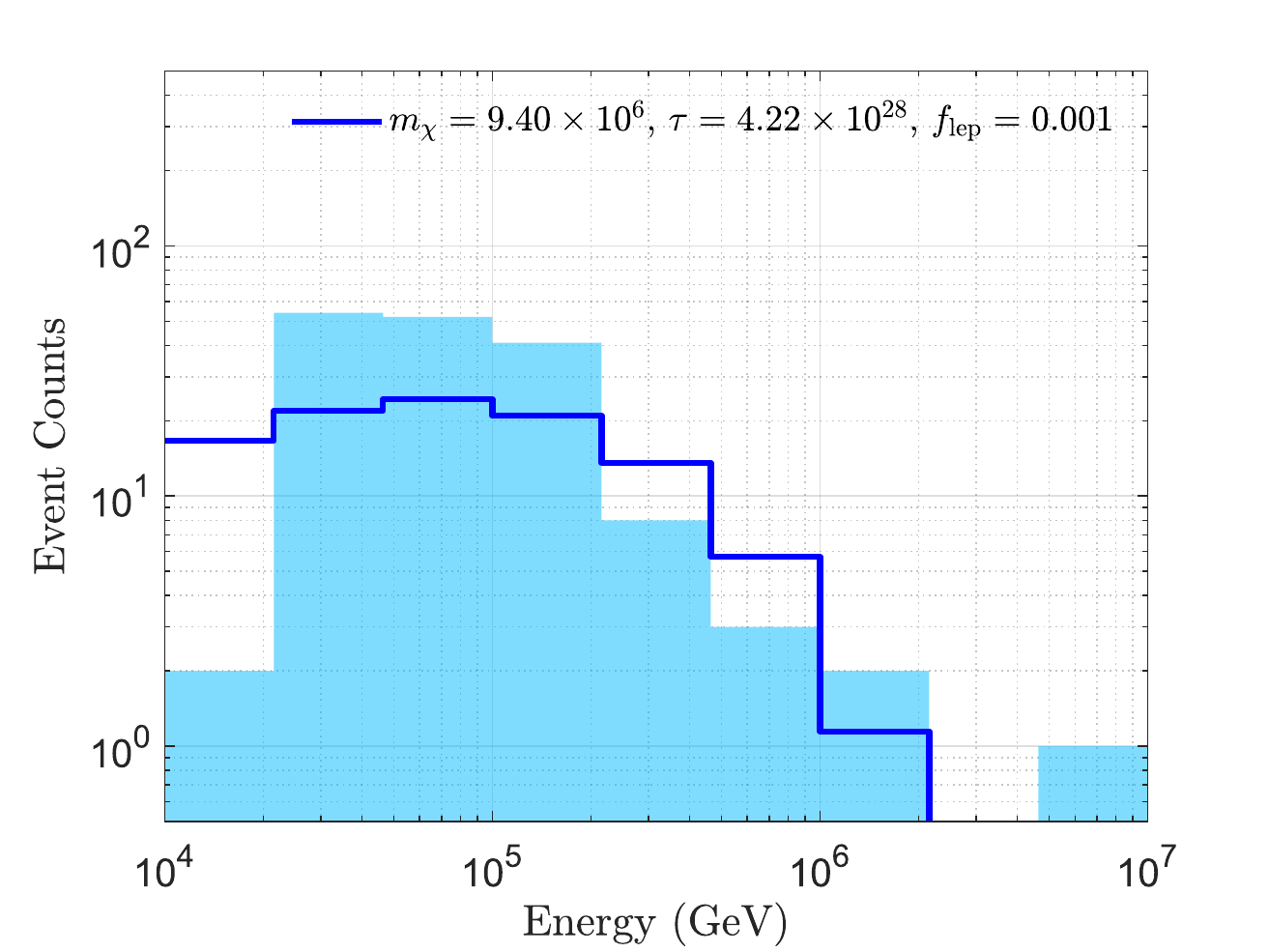}
    \caption{\label{fig:best_evnt} Predicted event counts from super-heavy dark matter decay compared with the IceCube observed event distribution. The shaded blue histogram represents the number of events derived from IceCube's 12-year dataset while the solid blue step curve shows the theoretical event distribution corresponding to the best-fit parameters.}
\end{figure}

Figure~\ref{fig:best_evnt} shows a comparison between the expected and observed number of neutrino events as a function of energy. The light blue shaded histogram displays the binned event counts from the 12-year IceCube High-Energy Starting Events (HESE) data. The solid blue step plot corresponds to the theoretically predicted number of events per energy bin, calculated using the best-fit values of the super-heavy dark matter mass $m_\chi$, decay lifetime $\tau$, and leptonic branching ratio $f_{\rm lep}$, obtained from the chi-square fitting described above.

\section{\label{section4} Discussions and Conclusions}
In this work we consider that superheavy dark matter from the early universe may undergo decay processes to produce neutrinos of energy range in and around PeV could explain the IceCube 12-year HESE data. Such a cascading decay process that proceeds through QCD cascades and electroweak cascade following leptonic and hadronic channels produce neutrinos that can explain the 12-year IceCube HESE data. A three parameter $\chi^2$ analysis has been made where the parameters are superheavy dark matter mass $m_\chi$, decay time $\tau$ and the contribution of the leptonic channel $f_{\rm lep}$ of the IceCube 12-year HESE data and the best fit values of these three parameters are obtained. In addition 1-$\sigma$, 2-$\sigma$, 3-$\sigma$ contours are also obtained for each of the three pairs of parameters chosen out of the three parameters. In doing so, the best fit values and 1-$\sigma$, 2-$\sigma$ and 3-$\sigma$ contours for each of the three pairs are obtained while marginalizing the third parameter. It has been found that the $\chi^2$ fit is satisfactory and hence it maybe said neutrino flux from the heavy dark matter decay could be a viable possibility for UHE HESE data at IceCube. The mass of such a heavy dark matter would be $\sim 9.4 \times 10^6$ Gev with decay lifetime $\sim 4.2 \times 10^28$ second. We also estimated in this work the contribution of the leptonic channel in the decay process that produces the neutrino flux. The best fit value of this fraction namely $f_{\rm lep}$ is found to be of the order of $\sim 10^{-3}$.

The overall agreement in both shape and normalization indicates that the SHDM decay model successfully reproduces the observed energy distribution of neutrino events. The consistency across multiple energy bins suggests that the chosen parameters provide a good fit to the data, supporting the plausibility of SHDM as a source of ultrahigh-energy neutrinos. Minor deviations between the model and the data may point to statistical fluctuations or hint at additional contributions beyond the SHDM decay scenario.

\bibliographystyle{plain}
\bibliography{ref}
\end{document}